# Excitonic Superconductivity in Charge Injected Organics

## M. P. Das and F. Green*

Department of Theoretical Physics, Institute of Advanced Studies,
The Australian National University, Canberra, ACT 0200, Australia

* Centre for Quantum Computer Technology, School of Physics, University of NSW,
Sydney, NSW 2052, Australia

**Abstract**

Transport of charge carriers can be controlled by doping through chemical and physical means. Unlike chemical doping, physical doping is carried out by a special technique through gate voltages in a field−effect transistor geometry. This technique keeps the carrier channels free from defects without complications from the crystalline structure and the dopant impurity sites. In this paper we discuss the occurrence of superconductivity at the interface of a device by an *unconventional* technique. We examine a dynamical pairing mechanism governed by the excitons in the active device. The pairing of charge carriers takes place when the system is in a nonequilibrium (driven) state. We discuss the physics of a plausible superconducting transition and suggest new experiments.

## 1. Introduction

In a provocative suggestion Fritz London raised the question whether it may be possible to have a superfluid state in a macro−molecular system. Inspired by this suggestion, William Little [1] proposed a new mechanism of molecular superconductivity. He envisaged how excitons, being bosons, could mediate the attractive interaction of two electrons on the Fermi surface, as do the phonons in the conventional BCS mechanism. Soon after, Ginzburg [2] extended this idea to interfacial superconductivity that can occur at the junction of a metal and a semiconductor system. Later, details of this idea were worked out in a model by Bardeen and coworkers [3]. In the simple weak−coupling picture the transition temperature can be obtained in analogy with a phonon−mediated superconductor as

$$T_c \sim \omega_{ex} \exp[-1/UN(E_F)],$$

where $\omega_{ex}$ is the energy of the mediator exciton, U is the electron−exciton interaction energy and $N(E_F)$ is the density of states at the Fermi level. As the scale of exciton energy is high, this model is capable of producing high temperature superconductivity.
Little's idea of excitonic superconductivity was pursued through the discovery of several organic superconductors. In 1979 the first organic superconductivity [4] was observed in $(TMTSF)_2PF_6$ (bis−tetramethyl−tetraselenafulvalene hexafluorophosphate). Over the past two decades a number of low−dimensional organic superconductors have become known with a recorded highest transition temperature above 10K. Three− dimensional "bucky balls" ($C_{60}$) with alkali doping are now also known to exhibit Tc above 40K. There has been ample discussion in the literature [5] regarding exciton–mediated superconductivity in CuCl. However, there is no definite conclusion whether excitons are the true mediators of pair formation within these materials.

In this paper we present a novel scenario wherein excitonic superconductivity is expected to occur for certain compelling reasons. In Sec. 2 we give a brief appraisal of the physics of organic field–effect transistors. The physical system we describe here is *not* in an equilibrium state. In Sec. 3 we review various instabilities against the normal metallic state, namely: charge–density, spin–density and superconducting states. Then in Sec. 4 we move on to a new type of phase transition occurring in nonequilibrium states and discuss how, in this case, the excitons may help to bind the Cooper pairs. We outline the formation of condensates. A number of experiments is suggested in Sec.5, to test the proposed superconductivity. We conclude with a brief summary.

## 2. Organic Field Effect Transistors

A transistor is an active electronic device, in which the carrier current has a directional property. A field effect transistor (FET) consists of a semiconductor/insulator material on which a conductive charge surface is created. Two ohmic contacts are made as source and drain electrodes. By using an insulating layer on the top of the charge layer a metal electrode is used to apply a gate voltage. This gate voltage controls the density of carriers on the two dimensional channel. More complicated gating (top gate and/or side gates) can be applied to control the width or thickness of the channel.

These same principles have been used for organic materials. Since the discovery of conducting polymers in the seventies, there has been plenty of activity on the organic FET idea. Though organic FETs have not been so efficient compared to the inorganic ones, there are many positive advantages in terms of purity, structural flexibility and low–temperature processing [6–9].

In semiconductors the bulk states are filled. Any interfaces in the system must have mobile charges if they have any excess charge at all. Charges due to dangling bonds will usually form a two–dimensional (2D) electron or hole gas. These charges are then manipulated by the gate potential to make the device operate, such that the channel can be conducting or insulating. A conducting or an insulating state of the channel corresponds to the 'on' or 'off' state of the device. A particular aim, at this point, is to make the device work as fast as possible for a useful switching action in integrated circuit structures.

Transport in organic 2D systems is substantially different from that in silicon or GaAs materials. Many organic materials are bonded by weak dispersion forces of van der Waals type. Because the van der Waals interaction is a physically generated force (no net charge transfer is involved), no chemical bond is cut when one makes a clean surface; therefore free charges are unavailable. One has to use high gate fields to induce an actual charge layer. Subsequently, these carriers aquire their mobility directly by the action of the source–drain potential. In that sense their transport properties are *extrinsic*; for this reason the mobility of organic transistors is quite low compared to the high intrinsic values in GaAs, or even in Si.

It has been pointed out [9] that the maximum room–temperature mobility of such a system is less than ~ 5 cm$^2$ V$^{-1}$ s$^{-1}$ and approaches a basic limit in the class of organic materials known today (see Fig.1 from ref. 9). At low temperatures, however, the mobility seems to be very high. It is this property that may be exploited for device making.

### 3. Possible instabilities in lower dimensions

By gate control of carriers in the induced charge layer, one can have a carrier density in the range of 109 to 1012 cm$^{-2}$. In the late seventies it was predicted that a noninteracting disordered electron system in spatial dimension d–2 cannot support a metallic state. Soon after it was reported that if one also takes into account the strong Coulomb interaction among the particles, the ground state can be metallic, not insulating. Later, in the mid–nineties, a metal-insulator transition (MIT) was experimentally observed in the 2D electron/hole gas on Si surfaces and in GaAs heterojunctions. A great deal of activity is currently under way to understand the physics of the metal–insulator transition in 2D systems [10].

Apart from the MIT, there are other possibilities for transitions to a variety of collective states, namely: charge–/spin–density wave states, glassy states with no long–range order, magnetic states, and superconducting states. Theorists consider these 'garden–variety' transitions within the ground state to be 'quantum phase transitions' and use different techniques to analyse them. For instance, they can study the occurrence of the lowest energy states of the system when an order parameter corresponding to a collective state is supported. These are mostly second–order phase transitions.

On the other hand, experimentalists measure a variety of properties corresponding to a particular state at finite temperature. For example, resistivity as a function of temperature is studied for superconductivity. One obtains a vanishing of the resistance below a certain critical temperature. This suggests a lower–energy state of the system for T<T$_c$, when the superconductng state wins out over the normal state.

For an interacting physical system, the free–energy difference between two competing states needs to be evaluated explicitly to decide which is the correct lowest–energy state. To carry out this, one needs an explicit model as well as a technique to study it. For the case of equilibrium phase transitions in lower dimension, see e.g. Ref.11.

### 4. Nonequilibrium excitonic superconductivity: a possible scenario

A two–dimensional electron/hole gas in an active transistor is always away from equilibrium. To understand the occurrence of superconductivity in the current carrying channel of the transistor, it may not be appropriate to use the conventional notions of BCS theory. Here we make a novel and plausible argument that superconductivity arises in the 2D system as a direct outcome of its nonequilibrium state.

An active transistor is in a dynamical steady state shared among electrons, holes, and their bound (excitonic) states. In the presence of the gate and source–drain (driving) fields, the excitonic bound states form between the excited electrons and the holes that they leave behind. Once such a gas of excitons is available, there is a possibility of excitonic condensation. In presence of a bias, however, exciton condensation is not favoured against other competing instabilities. Here we consider the superconducting state as an alternative to the excitonic condensation.

In the effective interaction experienced by the electron–hole charge pairs, the excitons act as mediators. They likewise intervene in the repulsive interactions among the similar charge carriers. The mediation of excitons generates a net attraction. Since the excitonic interaction is dynamic, the exciton dispersion has an important role to play.

In conventional BCS pairing theory, the binding takes place between same charge particles of equal and opposite momentum at the Fermi surface. For s–wave superconductors, the spins of the pairing particles opposite. Since the electronic system in the present problem is under the strong action of electric fields, the Fermi surface is displaced and also distorted. In making the pairs, there is no longer the opportunity to *exactly* match momentum states $k_F$ and $-k_F$ with opposite spins. Therefore, the pairing state will acquire a nontrivial total momentum (with the help of the exciton dispersion energy). This picture is similar to a model considered earlier by Fulde and Ferrel [12] and Larkin and Ovchinnikov[13] in presence of a weak magnetic field.

To formulate this problem we use the Gorkov–Keldysh nonequilibrium Green– function approach [14,15]. In this theory the dynamical energy gap is given by the Keldysh part of the Green function $\mathbf{g}_K$,

$$\Delta(r,\omega) \sim \text{Im } (\lambda_{ex}/1+\lambda_{ex}) \_ dE \; \mathbf{g}_K(E; r, \omega),$$

where $\lambda_{ex}$ is the electron–exciton coupling constant. This quantity is a parameter of the theory. However, from the knowledge of the dielectric function $\varepsilon(q,\omega)$ of the semiconductor material and by using the sum rules [3], we can estimate $\lambda ex$. For a given experimental condition (gate and bias voltage as well as temperature), we may calculate the energy gap and transition temperature. Details of this formalism are to be published elsewhere.

## 5. Suggested experiments

(a) *Materials*

For our dynamical mechanism to operate we need to materials with high dielectric polarizability. We recall that any effective interaction depends on the nature of the dielectric function $\varepsilon(q,\omega)$. The latter is given by the polarization function $\chi(q,\omega)$. Organic molecular systems, which have nonspherical orbitals, can have a particularly larger dielectric polarizability. A host of aromatic molecules is known to have such enhancements. Possible candidates of our device are: family of n–cenes (anthracene family), oligomers and conjugated polymers. Some of these materials also exhibits highly nonlinear polarizabilities (at higher electric fields).

(b) *Meissner effect with and without gate voltage*

The disappearance of resistance for $T<T_c$ is one test for the occurrence of superconductivity, but it is not the conclusive one. A crucial test of our proposal is the observation of the Meissner effect. A superconductor is a perfect diamagnet that expels all magnetic flux lines out of its interior material. If the superconductivity is induced by the dynamics of the charge carriers, then it could exhibit the Meissner effect for $T < T_c$, but *only* in the presence of a gate voltage. On switching the bias off, the superconductivity, and the Meissner effect, must disappear.

(c) *Optical and tunnelling spectroscopy to investigate the gap*
Once a noneqilibrium superconductor operates under bias, one can carry out optical and tunnelling spectroscopy to probe the nature of the superconducting energy gap. The superconducting gap can be studied as a function of gate and bias voltages. From the tunnelling I–V spectra, the electron–exciton coupling constant can be estimated.

## 6.Summary

Electronic technology is already well on the way to device engineering at near atomic scales. To retain control over the hugely increased switching rates that will then dominate device physics and design, we must look widely for novel material systems.
To replace the more traditional inorganic electronic materials like Si, Ge and GaAs, a new trend is emerging to use organic molecular solids for making novel field effect transistors. In this paper we have briefly discussed some novel aspects of organic transistors. At low temperatures the current–carrying active channel is likely to undergo one or more in a range of possible electronic phase transitions. We have considered superconductivity as a possible candidate, and we have discussed a fundamental mechanism by which superconductivity can be induced with the help of excitons. The novelty of our theoretical proposal is the occurrence of superconductivity in a *nonequilibrium* environment. We have suggested some organic molecular materials with high dielectric polarizabilty as candidates. We have also suggested crucial tests of our proposal through measurements of the Meissner effect, and of optical and tunnelling spectroscopy.